\documentclass[12pt]{article}
\begin{document}
\hoffset = -1 truecm
\voffset = -2 truecm
\baselineskip = 12pt
\input epsf

\title{\bf
Fermion Self-Energy and Chiral Symmetry Breaking from Four-Fermion and
Gauge Interactions\thanks{The project supported partially by National
Natural Science Foundation of China and by Grant No.LWTZ-1298 of the Chinese
Academy of Sciences.} 
}
\author{
{\bf
Bang-Rong Zhou}
\\
Department of Physics, The Graduate School at Beijing \\
University of Science and Technology of China, Academia Sinica \\
Beijing 100039, {\bf China}\thanks{Mailing address.}\\
and \\
CCAST ( World Laboratory ) P.O.Box 8730, Beijing 100080,
{\bf China}
}
\date {}
\newpage

\maketitle

\begin{abstract}
The exact analytic solutions of the linearized Schwinger-Dyson equation of 
fermion self-energy are used to obtain the effective four-fermion and gauge 
coupling criticality curves for dynamical chiral symmetry breaking. The 
results show that when the zero-momentum gauge coupling 
$\alpha(0) < \alpha_0(0)$, the critical gauge coupling in the pure gauge 
interaction case, the minimal critical four-fermion coupling 
$\beta_{\rm min}$  is always non-zero and positive and will go up as  the 
$\alpha(0)$ decreases. The use of the exact solutions also allow us to make 
quite definite estimations of the momentum scales where chiral symmetry  
breaking would happen if the values of an infrared parameter $\xi$ are given 
separately.
\end{abstract}

%%-main body of paper-%%

\indent It is an interesting problem to determine the critical coupling
constants and the momentum scales where chiral symmetry breaking will occour
in a theory with effective four-fermion and gauge interactions.  It has not
only fundamental theoretical significance but also can find its important
application in Technicolor(TC) [1], Extended Technicolor (ETC) [2] and the
Top-quark condensate theory [3-6] of dynamical electroweak symmetry breaking.
An efficient approach to persue this problem is to use the Schwinger-Dyson(S-D)
equation of fermion self-energy [7-13,4].  When effective four-fermion
interactions are included in, different from the case of pure gauge
interactions, one must introduce a finite momentum
cut-off $\Lambda$ and this will produce an important impact on results.  For
instance, because the four-fermion coupling constant is defined at the scale
$\Lambda$, the chiral symmetry breaking must much more depend on the
ultraviolet (UV) asymptoticality of theory. On the other hand, in order to
obtain some more quantitative predictions of the scales of chiral symmetry
breaking we need the exactest possible solutions of the S-D equation rather
than only the UV asymptotic form of these solutions. Fortunately, we have
obtained the exact analytic solutions of the S-D equation of fermion 
self-energy in some linearization approximation [14].  In this paper, we will 
use these solutions to discuss dynamical chiral symmetry breaking under a 
finite  momentm cut-off.\\
\indent The Lagrangian of the system to be delt with may be described by
$${\cal L}=-\frac{1}{4}F_{\mu \nu}^aF^{a\mu \nu}+
\bar{\psi}i{\gamma}^{\mu}(\partial_{\mu}+igA_{\mu}^a\lambda^a)\psi+
h[(\bar{\psi} \psi)(\bar{\psi}\psi)-(\bar{\psi}\gamma_5\psi)(\bar{\psi}\gamma_5
\psi)] \eqno(1)$$
\noindent where $\psi$ is the bare-massless fermion field which is assigned
in the representation $R_{\psi}$ of the color gauge group $G$ with dimention
$d(R_{\psi})$, $A_{\mu}^a$ and $F_{\mu \nu}^a$ are respectively the gauge
field and corresponding field strength tensor, $\lambda^a$ is the generator
of the gauge group $G$, $g$ and $h$ are respectively the coupling constant of
the gauge and the chirally invariant four-fermion interactions.  It has been
well known [11,4,14] that, the integral S-D equation of the
fermion self-energy $\Sigma(x)$ in the ladder approximation and in the Landau
gauge can be reduced to the differential equation
$$\omega(x)\Sigma^{\prime \prime}(x)+[\omega'(x)+1]\Sigma'(x)=-\frac{b}{\tau(x)}
\frac{\Sigma(x)}{[x+\Sigma^2(x)]}\eqno(2)$$
\noindent together with the IR boundary condition
$$\Sigma'(0)=-\frac{b}{2(\ln{\xi})\Sigma(0)}\eqno(3)$$
\noindent and the UV boundary condition
$${\left\{\left[1+\frac{\beta}{b}\tau(x)\right]\omega(x)\Sigma'(x)+
\Sigma(x)\right\}}_{x={\Lambda}^2}=0 \eqno(4)$$
\noindent where  $x=p^2$ is the squared Euclidean four-momentum and $\Lambda$
is the momentum cut-off.  The constant
$$b=\frac{3C_2(R_{\psi})}{16\pi^2\beta_0} \eqno(5)$$
\noindent with the eigenvalue $C_2(R_{\psi})$ of the squared Casimir operator
of the gauge group $G$ in the fermion field representation $R_{\psi}$  and the
coefficient of the $\beta$-function of the gauge coupling $g$ to one loop order
$$\beta_0=\left[11C_2(G)-\sum_f4T(R_f)N_f\right]/48\pi^2\eqno(6)$$
\noindent where the standard denotations in gauge theory have been used and
in the flavor sum $\sum_f$ in Eq.(6),  the $\psi$ fermion and  all the lighter
colored fermion flavors in the $G$-representation $R_{f}$ will be included in.
In the derivation of Eqs.(2)-(4) a continuous Ansatz [11] of the
running gauge coupling constant
$${\bar{g}}^2(x) = 1/\beta_0\tau(x) \eqno(7)$$
\noindent with
$$\tau(x) = \ln(\frac{x}{\mu^2}+\xi)\eqno(8)$$
has been used.  The scale parameter $\mu$ and the IR parameter $\xi$ are
optional except that $\xi>1$ is required so as to avoid the IR singularity
of ${\bar{g}}^2(x)$.  The constant $\beta$ is connected to the strength of
the four-fermion interactions and defined by
$$\beta=\frac{d(R_{\psi})h \Lambda^2}{2\pi^2}\eqno(9)$$
\noindent It is noticed that $\beta$ appears only in the UV boundary condition
(4), not in the equation (2).  The function $\omega(x)$ is defined by
$$\omega(x)={\left[\frac{1}{x}+\frac{1}{(x+\xi \mu^2)\tau(x)}\right]}^{-1}
\eqno(10)$$
\noindent The linearization approximation of Eq.(2) means that [14] \\
1) the $\Sigma^2(x)$ in the denominator of the right-handed side of Eq.(2)
will be replaced by the assumption
$$\Sigma^2(0)=\xi \mu^2; \eqno(11)$$
\noindent 2) the function $\omega(x)$ will be substituted by its approximate
expression
$$\omega(x)\simeq \frac{\tau(x)}{1+\tau(x)}(x+\xi \mu^2)\eqno(12)$$
\noindent so that Eq.(2) could be solved exactly.  The assumption (11) is
permissible since the parameters $\xi$ and $\mu$ are both undetermined
theoretically.  The approximation (12) is valid if
$$\frac{x}{\xi \mu^2}\left[1+\frac{1}{\ln{(\frac{x}{\mu^2}+\xi)}}\right] \gg 1
\eqno(13)$$
\noindent and it is certainly satisfied when $x\gg \xi \mu^2$. In the following
we will also extend the solutions of the linearized equation down to $x=\xi
\mu^2$ as a further approximation. \\
\indent In this way, the S-D equation (2) can be changed into that
$$\frac{\tau}{1+\tau}\Sigma^{\prime \prime }(\tau)+
\left[1+\frac{1}{{(1+\tau)}^2}\right]\Sigma'(\tau)+
b\frac{\Sigma(\tau)}{\tau}=0 \eqno(14)$$
\noindent Eq.(14) may have exact analytic solution with the general expression
$$\Sigma(\tau)=A\Sigma_{\rm irreg}(\tau)+B\Sigma_{\rm reg}(\tau) \eqno(15)$$
\noindent where $A$ and $B$ are two real constants and
$\Sigma_{\rm irreg}(\tau)$ and $\Sigma_{\rm reg}(\tau)$ are two linearily
independent solutions of Eq.(14)
$$\left. \begin{array}{l}
         \Sigma_{\rm irreg}(\tau) \\
         \Sigma_{\rm reg}(\tau)
         \end{array}\right\}=
  \left.\begin{array}{l}
         A_i \\
         A_r
        \end{array}\right\}M(\tau)+c.c. \eqno(16)$$
\noindent where
$$A_i=i(-1)^{\alpha}\frac{\Gamma(\bar{\alpha})}{\Gamma(\gamma)}
\frac{{\rm sin}(\pi \bar{\alpha})}{{\rm sinh}(2\pi \eta)} , \ \
A_r=-i\frac{\pi}{b{\rm  sinh}(2\pi \eta)|\Gamma(\bar{\alpha})|^2}
  \frac{\Gamma(\bar{\alpha})}{\Gamma(\gamma)} \eqno(17)$$
\noindent and
$$M(\tau)=e^{-\tau}{\tau}^{-\frac{1}{2}-i\eta}
 \left[\frac{\gamma}{2} \ {_1F_1}(\bar{\alpha}; \gamma; \tau)+
       \frac{\bar{\alpha}}{\gamma}\tau \ {_1F_1}(\bar{\alpha}+1;\gamma+1; \tau)
 \right] \eqno(18)$$
\noindent with the denotations
$$\begin{array}{l}
  \eta=\sqrt{b-\frac{1}{4}}, \\
  \alpha=-\frac{1}{2}+b-i\eta, \\
  \bar{\alpha}=\frac{1}{2}-b-i\eta, \\
  \gamma=1-i2\eta
  \end{array} \eqno(19)$$
\noindent and that $\Gamma(\bar{\alpha})$ is Gamma function and
${_1F_1}(\bar{\alpha};\gamma; \tau)$ is confluent hypergeometric (Kummer) 
function.  When
$\tau \rightarrow \infty$, the solutions (16) will approach their UV asymptotic
forms [9]:
$$\begin{array}{l}
\Sigma_{\rm irreg}(\tau)\stackrel{\tau \rightarrow \infty}{\rightarrow}
\tau^{-b} \\
\Sigma_{\rm reg}(\tau)\stackrel{\tau \rightarrow \infty}{\rightarrow} e^{-\tau}
{\tau}^{b-1}
\end{array} \eqno(20)$$
\noindent However, it is emphasized that in the case with the finite momentum
cut-off $\Lambda$ owing to presence of the four-fermion interactions, the
distinquishment between $\Sigma_{\rm irreg}(\tau)$ and $\Sigma_{\rm reg}(\tau)$
is not important and one must consider the general solution (15) as a linear
combination of the two independent solutions, i.e. normally both $A\neq 0$
and $B\neq 0$.  \\
\indent The solution (15) based on the approximation (12) is obviously
inapplicable in the region $x<\xi \mu^2$.  Therefore, we must impose the IR
boundary condition of the solution at $x=\xi \mu^2$ [14]
$${\left.\frac{d}{d\tau}\Sigma(\tau)=b{(\frac{d\tau}{dx})}^{-1}
\frac{d}{dx}\left[\frac{1}{x\tau(x)}\right]
\int_0^xdy\frac{y\Sigma(y)}{y+\Sigma^2(y)}\right|}_{x=\xi \mu^2} \eqno(21)$$
\noindent where the expression of $\Sigma(x)$ in $x<\xi \mu^2$ can be
approximated by the solution of Eq.(2) at small $x$, i.e.
$$\Sigma(x)=\Sigma(0)\left\{1-\frac{b}{2(\ln\xi ){\Sigma}^2(0)}x+
\frac{b}{6(\ln\xi ){\Sigma}^4(0)}\left(1-\frac{b}{2 \ln\xi}\right)x^2\right.$$
$$\left.-
\frac{b}{12 (\ln\xi ){\Sigma}^6(0)}\left[1-\frac{5b}{3\ln \xi}+
\frac{b^2}{3\ln^2\xi}\right]x^3+\cdots\right\}
\  \  \  {\rm for \ } x<\xi \mu^2 \eqno(22)$$
\noindent It is indicated that the solution (22) submits the IR boundary
condition (3) at $x=0$.  Considering the assumption (11) and the fact that
$\Sigma(x)$ must be continuous function at $x=\xi \mu^2$ we may change Eq.(21)
into that
$$\Sigma'(\tau_1)= - P(\tau_1)\Sigma(\tau_1),  \ \ \tau_1=\ln (2\xi) \eqno(23)$$
\noindent where
$$P(\tau_1)=b\frac{2\tau_1+1}{\tau_1^2}\frac{1}{f(1)}
\int_0^1dt\frac{tf(t)}{t+f^2(t)} \eqno(24)$$
$$f(t)=1-\frac{b}{2\ln \xi}t+
\frac{b}{6\ln \xi}\left(1-\frac{b}{2\ln \xi}\right)t^2-
\frac{b}{12\ln \xi}
\left(1-\frac{5 b}{3 \ln \xi}+\frac{b^2}{3\ln^2 \xi}\right)t^3
\eqno(25)$$
\noindent Now the whole problem becomes to seek the physical solution with
the form (15) which also satisfies the UV and the IR boundary conditions (4)
and (23).  The UV boundary condition (4) can be rewritten and turned into that
$$\beta=\frac{b}{\tau(\Lambda^2)}\left[-1-
\frac{1+\tau(\Lambda^2)}{\tau(\Lambda^2)}
\frac{\Sigma(\tau(\Lambda^2))}{\Sigma'(\tau(\Lambda^2))}
\right] \eqno(26)$$
\noindent  It is pointed out that in view of the expression (21) of
$\Sigma'(\tau)$, when taking off the gauge interactions i.e. setting $b=0$ we
will have $\Sigma(x)\rightarrow m={\rm constant}$ and may obtain from Eq.(26)
that
$$\beta=\Lambda^2/\int_0^{\Lambda^2}dy\frac{y}{y+m^2}=
1/\left[1-\frac{m^2}{\Lambda^2}\ln \frac{\Lambda^2+m^2}{m^2}\right] \eqno(27)$$
which is just the gap equation of the Nambu-Jona-Lasinio (NJL) model [15]
and now identical with the UV boundary condition.  Denoting the strength of the
running gauge coupling at the momentum cut-off scale $\Lambda$ by
$$\alpha^{(\Lambda)}\equiv \frac{\bar{g}^2(\Lambda^2)}{4\pi}=
\frac{4 \pi b}{3C_2(R_{\psi})\tau(\Lambda^2)} \eqno(28)$$
\noindent we may obtain from Eqs.(15), (16) and (26) that
$$\beta = \frac{b}{\tau}
{\left[-1-\frac{1+\tau}{\tau}
\frac{(A_i+\frac{B}{A}A_r)M(\tau)+c.c.}
     {(A_i+\frac{B}{A}A_r)M'(\tau)+c.c.}\right]}_
     {\tau =\frac{3C_2(R_{\psi}){\alpha}^{(\Lambda)}}{4\pi b}} \eqno(29)$$
\noindent From the IR boundary condition (23), it may be found out that the
real constant
$$\frac{B}{A}=-\frac{A_i[M'(\tau_1)+P(\tau_1)M(\tau_1)]+c.c.}
                    {A_r[M'(\tau_1)+P(\tau_1)M(\tau_1)]+c.c.} \eqno(30)$$
\noindent It depends upon the single infrared parameter $\xi$ when $b$
is given in a particular model.  The value of $\xi$ can be fixed by
phynomenology.  However, here we prefer some general discussions about the
results of its possible values.  Besides $\xi >1$, a further theoretical
constraint on $\xi$ may come from the requirement that $\Sigma(\tau)$, as
mass function of the fermion, must be positive-definite. In particular,
it must be so that
$$\Sigma(\tau=\tau_1) >0 \eqno(31)$$
\noindent which , by Eq.(22), will mean a new lower bound to be imposed on
$\xi$. \\
\indent Eq.(29) actually give the $\beta-\alpha^{(\Lambda)}$ criticality
curve of chiral symmetry breaking for a general theory with effective
four-fermion and gauge interactions.  To make a concrete insight, we will
apply it to a model based on one-generation of technifermions where we
identify the $\psi$-field with one of the technifermion "flavors", the
group $G$ with the technicolor gauge group $SU(4)$.  In this case the
constant
$$b=\frac{135}{224} \eqno(32)$$
\noindent if the eight "flavors" of the one generation of technifermions
are included in the calculation of $\beta_{0}$.  It follows from Eq.(31)
that the constraint on $\xi$ will be
$$\xi > 1.30017 \eqno(33)$$
\noindent we find that $B/A$ is sensitive to the change of $\xi$ and has
the following results:
$$\frac{B}{A}\left\{\begin{array}{l}
  <0, \\
  >0 \ ({\rm from} \ \infty \ {\rm to} \ 0),\\
  <0, \\
  \end{array}\right.
  {\rm if}\left\{
  \begin{array}{l}
  1.30017<\xi <1.44043 \\
  1.44043<\xi <3.09975 \\
  3.09975< \xi <\infty \\
  \end{array} \right.\eqno(34)$$
\noindent We can obtain the further constraint on $\xi$ from $\Sigma(\tau)>0$
for all $\tau >\tau_1$.  By using the continuity of $\Sigma(\tau)$ at
$\tau =\tau_1$, we get from Eqs.(15) and (16) that
$$\frac{\Sigma(\tau)}{\Sigma(\tau_1)}=
\frac{(A_i+\frac{B}{A}A_r)M(\tau)+c.c.}
     {(A_i+\frac{B}{A}A_r)M(\tau_1)+c.c.} \ \ \ \ {\rm if} \ \tau \geq \tau_1
     \eqno(35)$$
\noindent Since it has been demanded that $\Sigma(\tau_1)>0$, the sign of
$\Sigma(\tau)$ for $\tau >\tau_1$ will be determined by the sign of the
right-handed side of Eq.(35) which depends on $\xi$. It is not difficult to
verify that $\Sigma(\tau)$ will become negative for large enough $\tau$ if
$1.30017< \xi < 1.44043$.  For example, $\Sigma(\tau)<0$ if $\xi =1.35$ and
$\tau >1.6555$ and if $\xi =1.44$ and $\tau >7.92921$ respectively.  Hence
the corresponding $\Sigma(\tau)$ can not be physical solutions.  On the other
hand, if $\xi > 1.44043$ then we will have $\Sigma(\tau)>0$ and $\Sigma'(\tau)
<0$ for all $\tau \geq \tau_1$ and  the corresponding $\Sigma(\tau)$ could
be physical solutions. \\
\epsfxsize=7cm \epsfbox{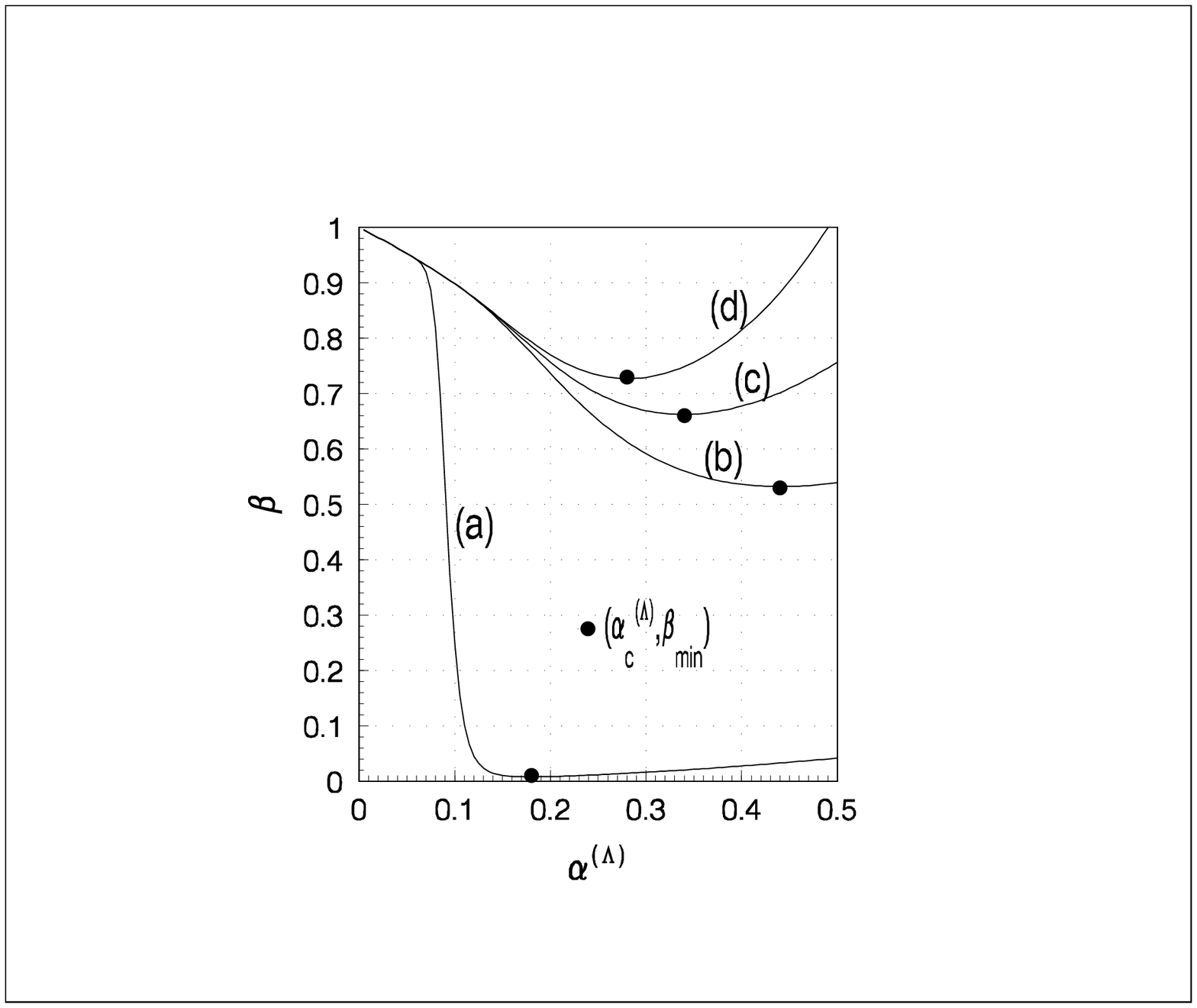}

Fig. 1.   The effective four-fermion and the gauge coupling $\beta$ and
$\alpha^{(\Lambda)}$ criticality curves are depicted for (a) $\xi=1.44044
(B/A\sim \infty)$; (b)$\xi=2 (B/A=4.12519)$; (c) $\xi=3.09975
(B/A=7.03327\times 10^{-6}\sim 0)$ and (d) $\xi=5 (B/A=-2.58869)$.
\\  \\
Table 1. The numerical values of relevant parameters to the curves (a), (b), (c)
and (d) in Fig. 1. \\
\begin{tabular}{ccccc}\hline
Curve & (a) & (b) & (c) & (d)   \\
\hline
$\xi$ & 1.44044 & 2 & 3.09975 & 5  \\
$B/A$  & 114865 & 4.12519 & $7.03327\times 10^{-6}$ & -2.58869  \\
$\alpha_1$ & 1.27247 & 0.97122 & 0.737967 & 0.584733 \\
$\alpha_c^{(\Lambda)}$ & 0.179016 & 0.440915 & 0.338492 & 0.280842  \\
$\beta_{\rm min}$ & $7.71772\times 10^{-3}$ & 0.532062 & 0.662304 & 0.726835 \\
$\Lambda_c \ (GeV)$ & --- & 1078 & 1745 & 2647 \\
\hline
\end{tabular} \\
\indent In Fig.1 we give the $\beta -\alpha^{(\Lambda)}$ curves for several
typical values of $\xi$.  The numerical values of relevant parameters to these
curves are listed in Table 1.  In Table 1 the denotation $\alpha_1=
4\pi b/3C_2(R_\psi)\tau_1$ has been used and the momentum scales $\Lambda_c$ are
found out from Eqs.(28), (8) and (32), $C_2(R_{\psi})=15/8$ and the values of
$\alpha_c^{(\Lambda)}$, if we assume the TC scale parameter $\mu=v=246 \ GeV$,
where $v$ is the vacuum expectation value of the Higgs field in the standard
electroweak theory.  For each $\beta -\alpha^{(\Lambda)}$ curve, we obtain a
$\beta$'s minimum point $({\alpha}_c^{(\Lambda)},{\beta}_{\rm min})$
in the UV region whose location has been indicated explicitly by a dot in
the curve.  It is seen from Table 1 that ${\alpha}_c^{(\Lambda)}$ are always 
much less than ${\alpha}_1$.  Since the solution (15) of Eq.(14) is reliable 
only in the region $x\gg \xi \mu^2$ i.e. $\tau \gg \tau_1$ or
${\alpha}^{(\Lambda)} \ll \alpha_1$, we will consider only the sectors of the
$\beta -{\alpha}^{(\Lambda)}$ curves in the regions
$\alpha^{(\Lambda)}\leq \alpha^{(\Lambda)}_c$ as physically acceptive
criticality curves of the coupling constants. \\
\indent The curve (a) shows that, when the value of $\xi$ is taken so that
$B/A\rightarrow \infty$, the minimum point of $\beta$ will approach the origin,
i.e. $(\alpha^{(\Lambda)},\beta)=(0,0)$ becomes a critical point.  This can be
verified rigorously by taking $B/A\rightarrow \infty$ and $\alpha^{(\Lambda)}
\rightarrow 0$ in Eq.(29). Since now $A\rightarrow 0$, we will leave only the
regular term $\Sigma_{\rm reg}(\tau)$  in Eq.(15)  corresponding to the
dynamical fermion mass generated by pure gauge interactions.  In this case no
finite momentum cut-off is needed because in an asymptotically-free gauge theory
the running gauge coupling $\alpha^{(\Lambda)}=\frac{\bar{g}^2(\Lambda)}{4\pi}
\rightarrow 0$ can be attained in the limit $\Lambda \rightarrow \infty$.  This
result is coincided with the conclusion in Ref.[4].  By means of the present
value $\xi=1.44044$ we may obtain the running gauge coupling at $p^2=0$ i.e.
$\alpha_0(0)=3\pi /7\ln \xi=3.689$, which, based on the analyses in Ref.[11],
can be regarded as the critical gauge coupling for occurance of chiral symmetry
breaking in the case of pure gauge interactions. \\
\indent The curve (b) shows the general feature of $\beta -\alpha^{(\Lambda)}$
in a finite momentum cut-off $\Lambda$.  Now we have both $A\neq 0$ and
$B\neq 0$, corresponding to existence of both the irregular term
$\Sigma_{\rm irreg}(\tau)$ and the regular term $\Sigma_{\rm reg}(\tau)$ in
$\Sigma(\tau)$.  We note that $(\alpha^{(\Lambda)}, \beta)=(0,1)$ is a critical
point which represents the case without the gauge interactions and is just the
NJL model.  Then the critical curve $\beta$ becomes a monotonically decreasing
function of $\alpha^{(\Lambda)}$ until $\alpha^{(\Lambda)}=\alpha^{(\Lambda)}_c
=0.440915$ where $\beta$ is down to its minimum value
$\beta_{\rm min}=0.532062$.  This means that for $\beta <1$, we must have
$\alpha^{(\Lambda)}>0$ in order to realize chiral symmetry breaking.  It may
also be concluded that if $\beta <\beta_{\rm min}$ then chiral symmetry breaking
could never happen.  We indicate that in present case, $\xi =2$ hence the
zero-momentum gauge coupling $\alpha(0)=3\pi/7\ln 2=1.9424$ is less than the
critical gauge coupling $\alpha_0(0)$ in tha case of pure gauge interactions and
this can explain why we need a non-zero lower bound $\beta_{\rm min}$ of the
four-fermion coupling $\beta$ so that chiral symmetry breaking could  occur only
at $\beta > \beta_{\rm min}$.  A numerical estimation of the scales of chiral
symmetry breaking can be made from the $\beta -\alpha^{(\Lambda)}$ criticality
curve in the region $\alpha^{(\Lambda)}\leq \alpha^{(\Lambda)}_c$.  It is
obtained from the third column in Table 1 that if $\beta > \beta_{\rm min}$,
then the corresponding scales of chiral symmetry breaking will be at
$\Lambda > \Lambda_c=1078 \ GeV$. \\
\indent The curve (c) shows another extreme case where $B/A\rightarrow 0$, i.e.
only the irregular term $\Sigma_{\rm irreg}(\tau)$ is left in $\Sigma(\tau)$.
The result remains to come from a combined effect of both the four-fermion and
the gauge interactions.  By comparing the curve (c) with the curve (b) we find
that both the $\beta -\alpha^{(\Lambda)}$ criticality curves have the similar
shape.  A numerical check indicates that as the value of $\xi$ increases
$\beta_{\rm min}$ will always go up.  This is plausible because the increase of
$\xi$ will imply the decrease of the corresponding $\alpha(0)$ , and this has to
be compensated by rising of $\beta_{\rm min}$ so that chiral symmetry breaking
could happen.  We also note that when $\xi$ increases the corresponding
$\alpha^{(\Lambda)}_c$ will arise at first but then come a continuative drop if
$\xi >1.528$.  This implies that for $\xi > 1.528$,  as $\xi$ increases we need
stronger four-fermion interactions but weaker gauge interactions so as to
realize dynamical chiral symmetry breaking. Consequently, a higher scale of
chiral symmetry breaking is expected.  This is shown in Table 1.  For example,
for $\xi =3.09975$ we need $\beta \geq \beta_{\rm min}=0.662304$ and  the
corresponding scales of chiral symmetry breaking will be
$\Lambda \geq \Lambda_c=1745 \ GeV$. \\
\indent The curve (d) shows another different case where $B/A <0$, i.e. the
irregular and the regular term of $\Sigma(\tau)$ have opposite sign but the
whole $\Sigma(\tau)$ remains to be positive.  In this case, the shape of the
curve is still similar to the ones of the curve (b) and the curve (c) except
that it has bigger $\beta_{\rm min}$ and smaller $\alpha^{(\Lambda)}_c$. Hence
higher scales of chiral symmetry breaking will be expected, e.g.
$\Lambda \geq \Lambda_c=2647 \ GeV$ for
$\beta \geq \beta_{\rm min}=0.726835$ . \\
\indent A complete determination of the $\beta- \alpha^{(\Lambda)}$ criticality
depend on the value of $B/A$ which is fixed by the IR boundary condition and
in present scheme by the value of $\xi$.  In addition, more exact consideration
of non-linearity of $\Sigma(\tau)$ in small $\tau$ region is also necessary.
These facts also imply that we can not make a completely precise prediction of
the scale of chiral symmetry breaking unless we have precisely known the
IR behavier of the gauge coupling constant and the non-linearity behavier of
$\Sigma(\tau)$ either from a fundamental theory or from phenomenology.
However, by means of the exact solutions of the linearized S-D equation of 
fermion self-energy,  we have given the general feature
of the $\beta- \alpha^{(\Lambda)}$ criticality curve which do not depend on the
details of the IR behavier of the theory.  In particular, we see from
Fig.1 that in the extreme UV region where ${\alpha}^{(\Lambda)} \approx 0$,
the four curves (a), (b), (c) and (d) with different values of the IR parameter
$\xi$ almost coincide with each other. On the other hand, once the IR 
parameter $\xi$ is given separately, we may make quite definite estimations of
the  scales of chiral symmetry breaking.  All these show obvious advantages of 
the approach of exact solutions over general numerical methods. \\
\indent The paralell analyses may be applied to the top-quark condensate scheme
of electroweak symmetry breaking if we identify $G=SU_c(3)$ and the 
$\psi$-field with the top-quark field. If the six quark flavors are included 
in calculation of ${\beta}_0$ we will have $b=4/7$. Similar results to the ones
in TC theory can be obtained except that the scale of chiral symmetry breaking 
would be much lower.
In the case of pure gauge interactions, i.e. $B/A \rightarrow \infty$, we have
$\xi =1.40025$ and the corresponding zero-momentum gauge coupling $\alpha(0)=
4\pi /7\ln \xi=4.9208$ which is bigger than the result from numerical analyses
[11] if it is identified with the critical gauge coupling for chiral
symmetry breaking.  The difference could be attributed to the method and
approximation used here.  To estimate the scale of chiral symmetry breaking, let
us consider the case of $\xi=2$. A similar $\beta - \alpha{(\Lambda)}$
criticality curve to the curve (b) in Fig.1 will be obtained but with $\beta =
\beta_{\rm min}=0.570637$ at ${\alpha}^{(\Lambda)}={\alpha}^{(\Lambda)}_c=
0.58046$ which corresponds to the smallest scale of chiral symmetry breaking
$\Lambda_c=408 \ GeV$ if we take the scale parameter $\mu$ to be the mass of
$Z$-boson $M_Z=91.187 \ GeV$ [16]. Since the resulting
$\Lambda_c \ll \Lambda_{top}$, the momentum cut-off in the bubble diagram of
the four-fermion interactions, which is up to $10^{15} \ GeV$ (the grand
unification scale) in the simplest top-quark condensate scheme [6] and
also at least above $5 \times 10^3 \ GeV$ even in an exteded version of the
scheme including the fourth generation of fermions [17], the chiral
symmetry breaking will happen actually at a point near
$(\alpha^{(\Lambda)},\beta)=(0,1)$ in the $\beta -\alpha^{(\Lambda)}$
criticality curve. This only reproduces the fact that the top-quark condensate
scheme is essentially a NJL model with some small corrections from the color
(and in addition, electroweak) interactions.  Certainly, by means of the
S-D equation of fermion self-energy and the linearization approximation
presented in this paper, we could deal with the corrections to the fermion mass
from the color gauge interactions. This approach could become an alternative 
one of the renormalization group analyses extensively used in pursuing this 
kind of problem [5,6,18,19]. \\

\end{document}